November 19, 2010

# GRAPHITE VS GRAPHENE: SCIENTIFIC BACKGROUND

Yakov Kopelevich[1] and Igor A. Luk'yanchuk[2]

[1] Instituto de Física "Gleb Wataghin", Universidade Estadual de Campinas, Unicamp 13083-859, Campinas, São Paulo, Brasil ; kopel@ifi.unicamp.br

[2] University of Picardie, Laboratory of Condensed Matter Physics, Amiens, 80039, France, and L. D. Landau Institute for Theoretical Physics, Moscow, Russia; lukyanc@ferroix.net

Nobel Prize in Physics 2010 was given for "*groundbreaking experiments regarding the two-dimensional material graphene."*  In fact, before *graphene* has been extracted from *graphite* and measured, some of its fundamental physical properties have already been experimentally uncovered in bulk graphite. In this Letter to the Nobel Committee we propose to include those findings in the Scientific Background.

To: The Nobel Committee,
Class for Physics of the Royal Swedish Academy of Sciences

Dears Members of the Nobel Committee,

It appears that our present letter has been written practically at the same time as the Letter by Dr. Walt de Heer to the Nobel Committee commented online by Nature magazine by November 18[th], 2010 (doi:10.1038/news.2010.620).

With all our deep respect to the Nobel Committee and its valuable work, we feel that important issues are missed in the Scientific Background on the Nobel Prize in Physics 2010. According to the Official Announcement, the prize has been awarded for "*groundbreaking experiments regarding the two-dimensional material graphene*". However, before *graphene* has been extracted from *graphite* and measured, fundamental properties of graphene layers have already been experimentally uncovered [1, 2]. The results obtained on graphite were the basis on which Manchester's [3] and Columbia [4] groups built their research. We do believe that it would be only fair to mention the original results [1,2] and hence to establish the actual course of events.

We agree with the comment by Dr. de Heer that all electrical transport measurements given in Science 2004 by Novoselov et al. [5] have been performed on graphite but not single layer graphene samples, and some of those results are not original. Thus, presented by Novoselov et al. data on the Quantum Hall Effect (QHE) coincide with the results previously reported for graphite [1]; see Fig. 1 and Ref. [6]. Although the original work [1] has been cited by Novoselov et al. in their on-line preprint [7]: "*This observation offers further support for recent speculations about a possible quantum Hall effect behaviour in graphite*", it has been omitted in the Science paper. Certainly, the original finding of QHE in graphite should be commented in the "Scientific Background".

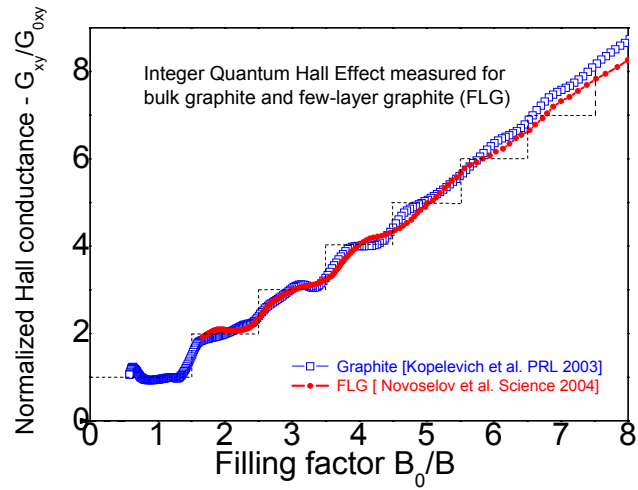

Fig. 1. Normalized Hall conductance $G_{xy} = 1/R_{xy}$ obtained for bulk highly oriented pyrolitic graphite (HOPG) [1] and few-layer-thick graphite samples [5]. The plot testifies a similar behavior of thick and thin graphite samples.

All the remarkable electrical properties of graphene described in "Scientific Background" are related to the unusual conic-like electronic spectrum of quasiparticles, known as Dirac Fermions, and the unambiguous evidence of Dirac fermions associated with decoupled graphene planes has been published one year before [2] of the articles by Novoselov [3] and Kim [4] in which we experimentally discovered the existence of the "*majority holes with a 2D Dirac-like spectrum*". Unfortunately in the Nature publication [3] this work was cited as the theoretical one. We believe that this key result obtained on graphite should also be mentioned in the "Background". More specifically, the quantum oscillation experiments [2] showed that Dirac fermions in *graphite* occupy an unexpectedly large phase volume, inconsistent with any previous 3D theoretical models. This is because the high-quality *graphite* contains nearly decoupled *graphene* planes whose electronic properties are governed by Dirac fermions. We stress that the occurrence of independent graphene layers with Dirac fermions in graphite has been confirmed in a large number of more recent experimental studies, see e. g. Refs. [8-11].

One consequence of the Dirac spectrum is that, the quantum effects are observable in *graphite* even at room temperature [6, 12] at low applied magnetic fields; $B < 1$ T. This is possible because of the exceptionally high mobility of electrons (holes) in *graphite ($10^6$ - $10^7$* cm$^2$/Vs). In fact, the quality of *graphene* planes in *graphite* is much higher as compared to separated (extracted) *graphene where the mobility is $10^5$* cm$^2$/Vs, at best. Hence, *graphite* and/or multilayer graphene is the most suitable material for quantum devices working under normal conditions, and can be considered as a natural solid state laboratory to test predictions of relativistic theories in the best way. In general, it has been demonstrated that few-layer graphite (e. g. grown on SiC substrate [13, 14]) is a more promising system for technological applications as compared to graphene.